# Elastic Models for the Non-Arrhenius Relaxation Time of Glass-Forming Liquids

Jeppe C. Dyre

*Department of Mathematics and Physics (IMFUFA),*
*"Glass and time" – DG Centre for Viscous Liquid Dynamics,*
*Roskilde University, Postbox 260, DK-Roskilde*
*Denmark*

**Abstract.** We first review the phenomenology of viscous liquids and the standard models used for explaining the non-Arrhenius average relaxation time. Then the focus is turned to the so-called elastic models, arguing that these models are all equivalent in the Einstein approximation (where the short-time elastic properties are all determined by just one effective, temperature-dependent force constant). We finally discuss the connection between the elastic models and two well-established research fields of condensed-matter physics: point defects in crystals and solid-state diffusion.

## INTRODUCTION

Viscous liquids and the glass transition is a classical subject, which has, however, only relatively recently attracted interest from mainstream physicists. It was originally cultivated in industry, in connection with polymer researches, and by (a few) chemists. But since 1990 the subject entered into condensed matter physics, and today it is an accepted discipline here. Part of the reason that physicists only late discovered the fundamental challenges of this research field may be that both experimental and theoretical studies of liquids for the last 50 years have been considered as parts of chemistry, not physics. This is perhaps a bit strange, because liquids – in particular viscous liquids approaching the glass transition – present important scientific challenges. Another reason that physicists ignored this field for so long may be that it was not generally recognized that highly viscous liquids have simple, common features which seem to hold for liquids held together by ionic, metallic, covalent, hydrogen chemical bonds. Thus there is a good chance that simple, universal models exist.

## THE THREE NON'S

Quite independently of their chemical nature viscous liquids may be generally characterized by the "three non's:"
1) **Non-exponential** time dependence of relaxations;
2) **Non-Arrhenius** temperature dependence of the average relaxation time;
3) **Non-linearity** of relaxations following large temperature jumps.

The last **non** is probably the least surprising, although a "large" temperature jump may, in fact, be rather small. As an example, a temperature jump from 208 K to 206 K of triphenyl phosphite makes the liquid relax to equilibrium much faster than the opposite jump from 204 K to 206 K. This is all well understood in terms of the fictive temperature, which controls relaxation together with temperature (in the simplest, phenomenological models), and we shall not dwell on this interesting phenomenon more here [1].

The non-exponential time dependence of relaxations is a general observation, which applies to both linear and non-linear relaxations. The former are usually studied in the frequency domain by means of, e.g., dielectric relaxation measurements or studies of the frequency-dependent specific heat, bulk or shear modulus. As is well known, these quantities are frequency dependent at much lower frequencies than in "ordinary" less viscous liquids. On approaching of the so-called calorimetric glass transition the Maxwell relaxation time becomes longer than 1 s, which induces loss peaks of the above-mentioned linear response functions typically at frequencies below 1 Hz. With modern electronics these measurements are fairly straightforward; it is even possible to go to frequencies corresponding to periods of 1 h or longer.

According to the fluctuation-dissipation theorem any linear-response function is the Laplace transform of an autocorrelation function (or time-derivative of such). For example, the dielectric loss is the Laplace transform of the time-derivative of the dipole autocorrelation function. If the latter is a simple exponential, the standard Debye linear response is predicted. This is virtually never observed; instead one finds loss peaks, which are asymmetric towards the high frequencies. A popular description of this is by means of the stretched exponential dipole autocorrelation function: $\exp(-(t/\tau)^\beta)$, which implies a loss peak at high frequencies varying with frequency as $\omega^{-\beta}$ whereas the low-frequency loss is predicted to (asymptotically) vary proportionally to frequency.

Our interest here is focused exclusively on the first non. This is the ubiquitous observation that the average relaxation time or viscosity – two quantities, which are roughly proportional in their temperature variations - only seldom follows the Arrhenius law (where the prefactor is a microscopic time of order 0.1 picoseconds):

$$\tau = \tau_0 \exp(\frac{\Delta E}{k_B T}). \qquad (1)$$

Almost all viscous liquids have a considerably more dramatic increase of the relaxation time upon cooling than predicted by this expression. If, however, one accepts Eq. (1) as expressing a well-established and basically correct prediction of rate theory (which derives from statistical mechanics), the activation energy $\Delta E$ must be temperature dependent. To explain the observed non-Arrhenius behavior the activation energy must increase when temperature is decreased. Typical data may be fitted if the log-log temperature derivative of the activation energy (its "index" [2]) numerically is between 2 and 6. This is the basic challenge we shall now address.

## THREE PHENOMENOLOGICAL MODELS

We define a phenomenological model as one predicting that the activation energy may be calculated from a macroscopic quantity. The elastic models to be considered in the next sections are phenomenological because they predict that the activation energy is proportional to short-time elastic constants of the liquid. First, however, we consider classical models based on *energy, volume, or entropy* as the controlling parameter.

Perhaps the most popular model of all is the Adam-Gibbs entropy model from 1965. This model predicts that the activation energy is inversely proportional to the configurational entropy. The model is generally regarded to give a good fit to experiment, where the configurational entropy (which cannot be measured) is usually estimated from excess entropy data, i.e., liquid minus crystal entropy at the same temperature. Nevertheless, the model does have a few problems: 1) The identification of excess and configurational entropy is problematic, because it assumes that the elastic properties of liquid and crystal are similar. If that were the case, the high-frequency sound velocity of liquid and crystal should have the same (weak) temperature dependence. This is not so; in viscous liquids the high-frequency sound velocity is much more temperature dependent than in crystals (or glasses). 2) When the model is compared to data one typically gets "cooperatively rearranging regions" containing only 4-8 molecules. This is not enough to justify the assumption going into the theory of regions, which relax independently of their surroundings. 3) The model predicts a second-order transition to a state of zero configurational entropy at a finite temperature, the so-called Kauzmann temperature. Such a state is unique and should have a simple description. None has been proposed, as far as known to the author. 4) Finally, the model is based on the *ad hoc* assumption that the activation energy is inversely proportional to the region volume.

The free-volume models, which exist in several non-equivalent versions, are all based on the assumption that the volume controls the activation energy. The free volume is that part of the liquid which is not occupied by the molecules. This is not a well-defined concept, unfortunately, but the basic idea is that molecules cannot move unless there is a certain amount of free volume present. In other words, a molecular rearrangement is possible only if the density locally (and briefly) is smaller than the average density. The model predicts that the activation energy over temperature is inversely proportional to the free volume. In other words, the relaxation time depends on temperature only because the liquid volume does so. If a liquid is put under high pressure and thus brought into a state of same volume as when cooled under ambient pressure, the free volume models predict that the relaxation time should be the same. This is contradicted by experiment, however [3].

Finally, we would like to mention the less famous models based on energy as the controlling variable. If $<E(T)>$ is the average energy of one rearranging region, in the simplest version the energy-controlled models predict that $\Delta E(T) = E_0 - <E(T)>$ [4,5], which obviously implies a non-Arrhenius behavior with activation energy increasing when temperature decreases. These models predict much too broad linear-response loss peaks, however, unless one is willing to accept very small region sizes (a problem similar to the entropy model's).

## ELASTIC MODELS

The history of these models is briefly reviewed in Ref. 6. The basic idea is that the short-time elastic properties, which are known to be much more temperature-dependent in viscous liquids, than in non-viscous liquids, glasses, or crystals, are responsible for the observed non-Arrhenius behavior. Interestingly, this idea is older than the entropy model. An early worker in this field was Mooney, who in 1957 suggested that - just as in free volume models - extra volume is needed locally for the molecules to be able to rearrange in a viscous liquid [7]. Referring to a flow event picturesque as a "quantum of liquid flow" he estimated the probability of a fluctuation creating a temporary, local density decrease by considering vibrations of the molecules around their equilibrium positions. This makes sense, because on the short time scale a viscous liquid is just like a (disordered) solid with molecules vibrating around potential energy minima. If these vibrations are expanded on a set of normal modes (phonons), Mooney calculated the probability that these by chance interfere to create the required density decrease. This resulted in an expression for the activation energy which is proportional to $mc^2$ where $m$ is the molecular mass and $c$ the longitudinal sound velocity. The latter quantity refers to the fixed structure of molecular positions, so in experiment it is to be identified with the high-frequency sound velocity, a quantity which is proportional to the high-frequency longitudinal modulus.

In Mooney's calculation only longitudinal sound waves play a role. This is because he argued that "concentrating free volume in one region happens only at the expense of producing higher density in other regions." That is not correct, because the total sample volume may fluctuate. Actually, if one adopts a simple model where a sphere via phonon interferences by chance expands its volume, the surroundings are not compressed at all. This was the basic observation behind the "shoving" model proposed in 1996 [8]. A sphere expanded in an elastic model induces radial displacements in its surroundings, which vary with distance from centre as $1/r^2$. This defines a radial vector field with zero divergence (compare to the Gauss law for a point charge) and thus no density changes. The result is that it is the *shear* modulus which is the relevant elastic constant. The shoving model predicts that, if $G_\infty$ is the so-called instantaneous shear modulus and $V_c$ a characteristic microscopic volume (of order the average molecular volume), the activation energy is given by

$$\Delta E(T) = V_c G_\infty(T). \qquad (2)$$

This expression has been compared to experiment on a number of molecular liquids, where it reproduces the non-Arrhenius behavior well [8].

These elastic models are mathematically equivalent if the instantaneous bulk and shear moduli are proportional in their temperature variations, because the longitudinal modulus is linear combination of the two former quantities. The correct bulk modulus refers to equilibrium fluctuations and is the isothermal, instantaneous bulk modulus, a quantity which is virtually impossible to measure. In an "Einstein" model of the short-time solid behavior of the liquid, all elastic properties are described via one effective, temperature-dependent force constant. Obviously, in this model the two above elastic models are equivalent.

We will not here review all elastic models, some of which refer to the vibrational mean-square displacement of a molecule and its temperature variation (the reader could consult, e.g., Ref. 6). The common feature is that the *vibrational, short-time* properties of the liquid determine the activation energy. This means that, e.g., properties which refer to the picosecond time scale are responsible for very long relaxation times in the second, hour or year range. How is that possible? The answer is that, while the long relaxation time of a viscous liquid is basically a measure of the time *between* two flow events involving the same molecule, the actual barrier transition is very fast and most likely takes place on the picosecond time scale. Thus it is entirely possible that liquid properties probed on this short time scale could determine the actual magnitude of the barrier.

There is no doubt that more experiments are needed before it is known how well elastic models in general explain the non-Arrhenius behavior of viscous liquids. On a qualitative level there is considerable encouragement to be found from the simple fact that the glass transition temperature scales with the melting temperature (the former is usually between 0.5 and 0.8 times the latter). This implies that there must be a Lindemann criterion (stating the melting takes place when the vibration amplitude of atoms is roughly 10% of the interatomic distance) also for the glass transition, and that the glass transition temperature should scale with the instantaneous moduli. This has been shown to be the case by Nemilov (who used the glass shear modulus as a measure of the instantaneous shear modulus of the glass-forming liquid just above the glass transition temperature) [9], and by Heuer and Spiess, who showed that the glass transition temperature is proportional to $mc^2$ [10].

## POINT DEFECTS IN CRYSTALS AND THEIR FORMATION AND MIGRATION ENERGIES

A classical subject of condensed matter physics is that of point defects in crystals. They play an obvious role in the understanding of how real-life crystal properties deviate from those of perfect single crystals. Of course, these defects are not the only important ones – metal properties, for instance, are dominated by dislocations (line defects).

There are of two sorts of point defects, vacancies and interstitials. A vacancy is simply a missing atom at a crystal site, while an interstitial is an extra atom squeezed in between the lattice sites. These two point defects are created in pairs in one and the same process, but afterward they are free to move away from one another and to the crystal surface, so there is no "conservation law" stating that the numbers of the two defects types are identical. Indeed, in typical metals the vacancy concentration is much larger than the interstitial concentration. Until recently, this led researchers to believe that interstitials in most cases play a minor role, but Granato has pointed out that their role is most likely much more important than this, because of the large vibrational entropy associated with an interstitial [11].

In thermal equilibrium the concentration of point defects is typically Arrhenius with an activation energy which is referred to as the "enthalpy of formation" (for simplicity we do not distinguish here between enthalpy and free energy, and shall refer to both simply as energy). It has been known for some time that the vacancy formation energies are proportional to the melting temperature [12]. It is also known that the melting temperature varies proportionally to the elastic moduli of a crystal, a statement which is equivalent to the above-mentioned Lindemann criterion.

In an interesting book from 1986 Varotsos and Alexopoulos collected data on point defects from a variety of solids [13]. These authors rationalized the data in terms of a simple model, according to which the formation energy of point defects (mainly vacancies) is equal to the bulk modulus of the crystal times the atomic volume (times a constant of order one). This means that point defect thermodynamics is intimately related to elastic properties of the crystal. Granato has used a similar idea, although he emphasizes that it is the shear modulus - and not the bulk modulus - which is important. In any case the Poisson ratio varies not very much (it is typically 0.3 for metals), implying a roughly universal ratio between the bulk and the shear moduli. Granato's theory involves an interesting feedback mechanism, namely that not only is the (interstitial) formation energy proportional to the shear modulus, but conversely the shear modulus itself is an exponential function of the (negative) interstitial concentration (and the shear modulus thus decreases upon heating towards the melting temperature).

Not only formation energy, but also the so-called migration energy of a point defect in a crystal is proportional to the elastic moduli. This brings us even closer to the basic flow event of viscous liquids, as we shall now see.

## CONNECTING THE ELASTIC MODELS WITH THEORIES OF POINT DEFECTS IN CRYSTALS

Almost nothing happens in a highly viscous liquid: In a computer simulation one sees repeated vibrations of the molecules around the fixed, somewhat random positions. Only seldom does a flow event take place, taking the molecules from one potential energy minimum to another. On the other hand, it is these rare flow events that are responsible for flow and which make it a liquid. In the present author's opinion a viscous liquid is more to be regarded as a "solid which flows" than as an ordinary (less-viscous) liquid.

Flow events are rare simply because a large energy barrier must to be overcome for the molecules to rearrange. Consider one such transition between two minima. Each of the minima corresponds to a solid in a sense, and the two "solids" before and after the flow event differ just by having a few molecules change positions (plus some additional minor elastic adjustments in the surroundings). The process taking the molecules from the initial minimum to the barrier may likewise be considered as a solid-to-solid transition, because the barrier (when optimized as happens increasingly at low temperatures) itself corresponds to a state of mechanical equilibrium - albeit an unstable such. The result is the following: The barrier energy is equal to the energy difference of two solid states, that differ by just having a few molecules change positions (plus the derived minor adjustments of the surrounding molecule positions). This is more or less the definition of a point defect, if one defines it in general terms without reference to a crystal lattice.

It is now clear that the elastic models embody old and well-established concepts of condensed matter physics by the basic postulate that the flow event activation energy is proportional to the elastic moduli of the fixed structure, the instantaneous moduli. The only new thing is that these quantities are surprisingly temperature dependent in viscous liquids, much more so than in less-viscous liquids like ambient water, in crystals, or in glasses.

## OUTLOOK

The "standard" phenomenological models for the non-Arrhenius behavior assume that the activation energy is controlled by either entropy or free volume (or energy in a less widely used model category). Both the entropy model and the free volume model (in the Cohen-Turnbull version) predict that an underlying phase transition to a state of infinite relaxation time (at a finite temperature) is the real cause of the slowing down of a viscous liquid as it is cooled and approaches the glass transition. There is no such underlying phase transition in the elastic models.

If the elastic models are correct, the liquid-glass transition has little in common with the ergodic-nonergodic transitions of mode coupling theory or of various spin and lattice models, and it also has little in common with the jamming transition of granular media. On the other hand, if correct, elastic models may be useful in practice, because measurements on the short time scale (e.g., via ultrasonics) could make it possible to directly monitor the structural relaxation time of a glass-forming melt just below the glass transition. That could be potentially quite useful in many applications.

## ACKNOWLEDGMENTS

This work was supported by the Danish National Research Foundation.

## REFERENCES

1. S. Brawer, *Relaxation in viscous liquids and glasses*, Columbus (OH): American Ceramic Society, 1985.
2. J. C. Dyre and N. B. Olsen, *Phys. Rev. E* **69**, Art. No. 042501 (2004).
3. C. Alba-Simionesco, *C. R. Acad. Sci. Paris (Ser. IV)* **2**, 203-216 (2001).
4. M. Goldstein, *Faraday Symp. Chem. Soc.* **6**, 7-13 (1972).
5. J. C. Dyre, *Phys. Rev. B* **51**, 12276-12294 (1995).
6. J. C. Dyre, T. Christensen, and N. B. Olsen, *cond-mat/*0411334 (2004).
7. M. Mooney, *Trans. Soc. Rheol.* **1**, 63-94 (1957).
8. J. C. Dyre, N. B. Olsen, and T. Christensen, *Phys. Rev. B* **53**, 2171-2174 (1996).
9. S. V. Nemilov, *Sov. J. Glass Phys. Chem.* **18**, 1-12 (1992).
10. A. Heuer and H. W. Spiess, *J. Non-Cryst. Solids* **176**, 294-298 (1994).
11. A. V. Granato, *Phys. Rev. Lett.* **68**, 974-977 (1992).
12. M. Doyama and J. S. Koehler, *Acta Metall.* **24**, 871-879 (1976).
13. P. A. Varotsos and K. D. Alexopoulos, *Thermodynamics of point defects and their relations with bulk properties*, Amsterdam: North-Holland, 1986.